\documentclass[epj]{svjour}
\usepackage{amsmath}
\usepackage{graphicx}
\usepackage{subfig}

\newcommand{\vk}{\ensuremath{\mathbf{k}}}

\newcommand{\vp}{\ensuremath{\mathbf{p}}}

\begin{document}

\title{Coboson formalism for Cooper pairs used to derive Richardson's equations}
\author{Monique Combescot\inst{1,2} \and Guojun Zhu\inst{2}}
\institute{Institut des NanoSciences de Paris, Universite Pierre et Marie Curie, CNRS, Campus Boucicaut, 140 rue de Lourmel, 75015 Paris, France \and Department of Physics, University of Illinois at Urbana-Champaign, 1110 W Green St, Urbana, IL, 61801}

\abstract{We propose a many-body formalism for Cooper pairs  which has similarities  to the one we recently developed for  composite boson excitons (coboson in short). Its Shiva diagram representation evidences that  $N$ Cooper pairs differ from $N$ single pairs through electron exchange only: no direct coupling exists due to the very peculiar form of the BCS potential. As a first application, we here use this formalism to derive Richardson's equations for the exact eigenstates of $N$ Cooper pairs. This gives hints on
why the $N(N-1)$ dependence of the $N$-pair ground state energy we recently obtained by solving Richardson's equations analytically in the low density limit, stays valid up to the dense regime, no higher order dependence exists even under large overlap, a surprising result hard to accept at first. We also briefly question the BCS wave
function ansatz compared to Richardson's exact form, in the light of our understanding of coboson many-body effects. }
\date{\today}
\maketitle

\section{Introduction}
It is commonly accepted that the Pauli exclusion principle
plays a key role in superconductivity. None the less, the
precise way Pauli blocking transforms a collection of single Cooper pairs into a superconducting
condensate, still is an open problem. This precise understanding goes through
the study of Cooper pairs not within the grand canonical ensemble as done in the
standard theory proposed by Bardeen, Cooper and Schrieffer (BCS), but within the canonical ensemble. To handle the
Pauli exclusion principle between a fixed number of interacting fermions is however  known  to be quite difficult when these fermions are paired. 
Turning to the grand canonical ensemble makes the task far easier. This is why superconductivity has been tackled this way, the two procedures being equally valid in the thermodynamical limit. Yet, adding fermion pairs one by
one constitutes the one and only way to fully control the increasing effect of Pauli
blocking from the dilute to the dense regime of pairs.

Five years after the BCS milestone paper\cite{BCS} on superconductivity, Richardson succeeded to solve this N-body problem formally\cite{Richardson1,Richardson2,Richardson1968,Richardson3} (see also reference \cite{gaudin,Ushveridze}) . He showed that 
the exact eigenstates of the Schr\"{o}dinger equation for an arbitrary number $N$ of pairs can be expressed in terms
of $N$ parameters, $R_{1}$,... $R_{N}$ which are solutions of $N$ coupled
non-linear equations, the energy of these $N$ pairs reading as $E
_{N}=R_{1}+...+R_{N}$. Although this exact form is definitely very smart, to
use it in practice is not that easy: Except in the infinite $N$ limit for which the BCS energy has been recovered \cite{Richardson3}, the equations giving $R_{1}$,... $R_{N}$  have not been, up to now, analytically solved for arbitrary $N$ and arbitrary interaction strength. The only approaches are numerical\cite{Duk,ortiz,delft}. This probably is why Richardson's solution has not had so far the
attention it deserves. Nowadays, these equations 
are commonly addressed numerically to study superconducting granules having small number of pairs\cite{Duk,delft2001,sierra2000,schechter2001}. 

Last year, we decided to tackle again these Richardson's equations because we wanted to reveal the deep connection which has to exist between two well-known problems, namely the
one-pair problem solved by Cooper and the many-pair problem considered by Bardeen, Cooper and Schrieffer. 
These two problems have intrinsic similarities: In
both cases, there is a ``frozen'' core of non-interacting electrons. Above this core, there is
a potential layer with attraction between up and
down spin electrons having opposite momenta. In the one-pair problem, this layer contains one electron pair 
only, while in the standard BCS configuration, the potential
layer is half-filled - a symmetrical potential 
on both sides of the Fermi level just corresponds to fill half the layer.
It is clear that, by adding more and more pairs to the frozen Fermi sea, we must go from the one-pair problem studied by Cooper to the dense regime studied by Bardeen, Cooper and Schrieffer. 

Although, at the present time, such a continuous pair increase
does not seem easy to experimentally achieve, this increase can at least be seen as a 
gedanken experiment to study the evolution of the energy spectrum when
the filling of the potential layer is changed, in order to understand the exact role of the Pauli
exclusion principle in superconductivity. 
This procedure can also be
seen as a simple but well-defined toy model to tackle the BEC-BCS crossover
since, by changing the number of pairs, we change their overlap. 
An overlap change has already been considered by Eagles \cite{Eagle}, 
and also by Leggett \cite{LeggettCrossover}, through the change of the interaction strength between pairs. In their approach, the number of pairs is fixed, so that Pauli blocking does not change when the overlap changes while it increases when the pair number increases. Consequently, the two types of overlap change do not involve the same physics. This is why to change the overlap by changing the pair number at constant potential is a fully relevant problem, complementary to the one studied in the past by Eagles and by Leggett. We wish to mention that Ortiz and Dukelsky have applied Richardson's approach to the BEC-BCS crossover problem with a different perspective, the overlap being varied by changing the interaction strength\cite{crossoverRich}, along Eagles' and  Leggett's idea.  Some interesting comparisons between the BCS ansatz and Richardson's solution also follow from their work.

Since the Richardson's procedure allows one to fix the pair number and vary 
this number at will from one to half filling, we seriously reconsidered 
solving these equations analytically in order to better understand how superconductivity develops from a collection of single pairs through the $N$ dependence of their ground state energy. By turning to the dimensionless form of Richardson's equations, we succeeded to solve these equations analytically at lowest order of density in the dilute limit on the single pair scale\cite{paper1}. 
Indeed, these equations do have a small dimensionless parameter which is the inverse of the number of pairs $N_{c}$ from which overlap between noninteracting single pairs would start. 
This allowed us to demonstrate that, for $N$ arbitrary large but $N/N_{c}$ still small, 
the energy of $N$ Cooper pairs reads, at lowest order in $1/ N_{c}$, as
\begin{equation}
E_{N}= N\left[ \left( 2\epsilon _{F_{0}}+\frac{N-1}{\rho _{0}}%
\right)-\epsilon _{c}\left( 1-\frac{N-1}{N_{\Omega }}\right) \right]
\label{eq:eN}
\end{equation}%
$\epsilon _{F_{0}}$ is the Fermi level energy of the frozen sea.  \\$\epsilon _{c}\approx
2\Omega \exp \left( -2/\rho _{0}V\right) $ is the single pair binding
energy for a small potential amplitude $V$ (weak-coupling limit). $\rho
_{0} $ is the density of states within the potential
layer, taken as constant. It linearly increases with sample size. $N_{c }=\rho _{0}\epsilon _{c} $ while $N_{\Omega }=\rho _{0}\Omega $ is the number of free pair states in the potential
layer, $\Omega $ being the layer extension.

A $N(N-1)$ dependence in the energy of $N$ pairs suggests interaction treated at lowest order in density. In spite of it, the above result fully agrees with the textbook energy obtained in the dense BCS configuration. In other words, all higher order terms in the $N$ dependence of the energy cancel exactly, even under strong overlap.  Indeed, the first term of Eq.\eqref{eq:eN} is
the exact energy of $N$ pairs in the normal state whatever this pair is. For a constant density of states $\rho _{0}$, the kinetic energy of $N$ free pairs above the frozen Fermi sea, is given by
\begin{multline}
{E}_{N}^{\text{(}normal)}= \\
2\left[\epsilon _{F_{0}}+\left( \epsilon _{F_{0}}+1/\rho _{0}\right) +\cdots
\;+\left( \epsilon _{F_{0}}+(N-1)/\rho _{0}\right)\right]
\end{multline}%
which is exactly equal to the first term of Eq. \eqref{eq:eN}.

If we now turn to the condensation energy in the BCS configuration, obtained, for a number of pairs corresponding to fill half the potential layer, we find,  according to Eq.\eqref{eq:eN}  
\begin{equation}
{E}_{N}^{\text{(}normal)}-{E}_{N}=\frac{N_{\Omega }}{2}\frac{%
\epsilon _{c}}{2}=\frac{1}{2}\rho _{0}\Omega ^{2}e^{-2/\rho _{0}V}
\end{equation}%
This result exactly matches the energy  $\rho _{0}\Delta ^{2}/2$ obtained by Bardeen, Cooper, Schrieffer within the grand canonical
ensemble using their wave function ansatz, since the gap $\Delta $ reads as $%
2\omega _{c}\exp \left( -1/\rho _{0}V\right) $ where $2\omega _{c}$ is just the
potential layer extension $\Omega $. 

The validity of a $N(N-1)$ interaction term over the whole density range seems to indicate that either Cooper pairs are not involved in many-body effects higher than 2x2, or some magic cancellation takes place even in the dilute limit on the single pair scale. This strongly indicates that some unrevealed physics must hide behind such a  surprising $N$-dependence, hard to accept at first. 

As the Pauli exclusion principle is said to play a key role in superconductivity while Pauli blocking between $N$ paired fermions is commonly known to be difficult to handle properly, it can appear of interest to approach many-body effects with Cooper pairs through a composite boson formalism similar to the one we have successfully developed for the many-body physics of excitons in semiconductors\cite
{CobosonPhysicsReports}.

The main purpose of the present work is to settle such a formalism. The physics being fully determined by the Hamiltonian, the major difference between an exciton gas and a set of Cooper pairs of course lies in the potential. Excitons interact via the Coulomb potential between its carriers. For Cooper pairs, we here take the usual ``reduced" BCS potential without questioning it. While its relevance has been proved in many physical effects, its main advantage surely is its simplicity which makes it ``solvable'' - even more than originally thought by Bardeen, Cooper and Schrieffer, as seen from Richardson's works and the quite recent analytical solution we found to his equations.
 
  Coulomb potential is long-range, the BCS potential, taken as separable, is short range. This can be a reason for Cooper pairs to stay bound at large densities while excitons break through a Mott transition when the density increases. However, to our opinion, the crucial difference between excitons and Cooper pairs lies in the fact that usual excitons  are made of fermion pairs having two degrees of freedom while the electrons which interact by the BCS potential have opposite momenta, so that electron pairs have one degree of freedom only.  Actually, there also are excitons made of pairs with one degree of freedom: Frenkel excitons. Those exist in organic materials while Wannier excitons are found in inorganic materials. The latter are made from a free electron and a free hole, attracted by intraband Coulomb processes. By contrast, Frenkel excitons are made of atomic excitations on ion sites which are delocalized into exciton by Coulomb processes between atomic levels.  It is worth noting that we have found the same  $N(N-1)$ dependence for the hamiltonian mean value taken between $N$ ground state Frenkel excitons \cite{frenkel}, while Wannier excitons have been shown to have terms in $N(N-1)(N-2)$ and higher\cite{monicOdil}.

 Another important difference between excitons and \\Cooper pairs is that Cooper pairs are said to stay bound in the dense regime, i.e., under strong overlap, while excitons dissociate through a Mott transition. As a result, excitons when they exist, always are in the dilute limit while the relevant regime for superconductivity is the dense regime. Due to this, creation operators for single exciton eigenstates are relevant operators to tackle the exciton many-body physics while the single Cooper pair operator is probably not a relevant operator in the dense regime. We can nevertheless develop a composite boson many-body formalism not for correlated pairs but for the free pairs out of which the BCS condensate is made. This is what we here do.

As a first interesting outcome of this formalism, we clearly see that, due to the very peculiar form of the reduced BCS potential, two pairs of free electrons with opposite spins and opposite momenta have an interaction scattering which is a succession of a fermion exchange between pairs followed by a fermion interaction \textit{inside} one pair. Since fermion exchange physically comes from the Pauli exclusion principle, this formalism evidences that two electron pairs interact, within the BCS potential, due to Pauli blocking only. However, as the Pauli exclusion principle acts between any number of pairs, exchange interaction scatterings - which originate from this Pauli exclusion - should a priori exist between more than two pairs. This strongly questions the $N(N-1)$ dependence of the $N$-pair energy we found: Why Pauli-induced $N$x$N$ exchanges do not show up through higher order terms in the energy?

The coboson formalism is capable of exactly handling Pauli blocking between an arbitrary number of composite bosons. Since the exact eigenstates of $N$ pairs have been shown to follow from Richardson's equations, a relevant first application of this formalism  is to address to the eigenenergies of $N$ pairs in order to show how  Richardson's equations follow from this formalism, and to possibly understand the origin of the eigenenergy $N$ dependency we found. In doing so, we see that $N-2$ pairs stay unchanged when the Hamiltonian $H$ acts on $N$ pairs. Since these pairs have one degree of freedom only, they cannot exchange their fermions in order to generate higeher order exchange Coulomb scattering as in the case of Wannier excitons. This can be the physical reason for the ground state energy of $N$ pairs to depend on $N$ as $N(N-1)$ only, with no higer order term, whatever $N$, a result hard to accept, especially when pairs strongly overlap.

The paper is organized as follow:

In section \ref{sec:beta}, we present the coboson formalism appropriate to many-body effects between the free
electron pairs on which Cooper pairs are constructed. We derive the Pauli and interaction scatterings for these free pairs.

In section 
\ref{sec:rich}, we use this formalism to rederive Richardson's form of the
exact eigenstates for $N=1,2,3,\cdots$ pairs interacting through the reduced
BCS potential, in order to see how the solution for general $N$ develops. We then derive this general $N$ solution explicitly.

In section III,  we physically analyze the role of the Pauli exclusion principle in a collection of Cooper pairs. We, in particular, show that the $1/(R_i-R_j)$ terms in Richardson's equations readily follow
from Pauli scatterings for fermion exchanges 
between electron pairs. The Richardson's parameters $R_i$  do have $N$ \emph{different} values just because 
of Pauli blocking between Cooper pair components. As a direct consequence, the $N$-pair ground state must be fundamentally different from the BCS ansatz. We then question this ansatz in the light of our general understanding of the many-body physics of composite bosons. In this section, we also briefly discuss the major difference between Cooper pairs and Wannier excitons which could possibly explain why the energy of $N$ Wannier excitons has terms in $N(N-1)(N-2)$ and higher while they do not exist for Cooper pairs.

In the last section, we conclude.

\section{Commutation technique for free fermion pairs making Cooper pairs\label{sec:beta}}

In our recent works on the many-body physics of composite bosons - essentially concentrated on semiconductor excitons - we have
proposed a ``commutation technique'' which allows an exact treatment of Pauli blocking between the fermionic components of these composite
bosons (cobosons in short). They appear through dimensionless ``Pauli scatterings'' which describe fermion
exchanges in the absence of fermion interaction. These dimensionless
scatterings, when mixed with energy-like scatterings coming from interactions
between the coboson fermionic components, allow us to deal with fermion exchanges
between any number of composite particles in an exact way. For a review on
this formalism and its applications to the many-body physics of
semiconductor excitons, see Refs. \cite%
{CobosonPhysicsReports}.

We here construct a similar formalism for the free electron pairs on which Cooper pairs are made.

\subsection{Exchange scattering}

We consider free fermion pairs with zero total
momentum 
\begin{equation}
\beta^{\dagger}_\vk=a^{\dagger}_{\mathbf{k} }b^{\dagger}_{-\mathbf{k} }
\end{equation}
 In the case of Cooper pairs, $a^{\dagger}_{\mathbf{k} }$ creates a  up-spin electron with momentum $\mathbf{k}$ while $b^{\dagger}_{\mathbf{-k} }$ creates a down-spin electron with momentum $\mathbf{-k}$. These $\beta^{\dagger}_\vk$ pairs have one degree of freedom
only, namely $\mathbf{k}$. This has to be  contrasted to the
most general fermion pairs $a^{\dagger}_{\mathbf{k} _1}b^{\dagger}_{\mathbf{k%
} _2}$, such as Wannier exciton pairs, which have two. $\beta^{\dagger}_\vk$ pairs actually have  some similarity to Frenkel exciton pairs \cite{frenkel}, the index $\mathbf{k}$ being then replaced by the excited ion site $n$. 

It is straightforward to show that the creation operators of these free fermion pairs commute 
\begin{equation}  \label{eq:bCom}
\left[\beta^{\dagger}_{\mathbf{k} ^{\prime}},\beta^{\dagger}_{\mathbf{k} }%
\right]  =0
\end{equation}
These free pairs thus are boson-like  particles.  It however is worth noting that while ${(a^{\dagger}_{\mathbf{k}})} ^2=0$
simply follows from the anticommutation of $a^{\dagger}_{\mathbf{k} }$
operators, the cancellation of ${(\beta^{\dagger}_{\mathbf{k}})} ^2$ does not follow from Eq.\eqref{eq:bCom}, but from the fact that ${(\beta^{\dagger}_{\mathbf{k}})} ^2$  contains ${(a^{\dagger}_{\mathbf{k}})} ^2$. The ${(\beta^{\dagger}_{\mathbf{k}})} ^2$  cancellation which comes from Pauli blocking, may appear  to be lost
when working with pair operators instead of  single fermion operators. We will see that this Pauli blocking is yet preserved in the commutation algebra for
free fermion pairs we develop.

If we now turn to creation and annihilation operators, their commutator reads
\begin{equation}  \label{eq:betacom}
\left[\beta_{\mathbf{k} ^{\prime}},\beta^{\dagger}_{\mathbf{k} }\right] 
=\delta_{\mathbf{k} ^{\prime}\mathbf{k} }-\mathit{D} _{\mathbf{k} ^{\prime}%
\mathbf{k} }
\end{equation}
where the ``deviation-from-boson operator'' of two zero-momentum free fermion pairs $\mathit{D} _{\mathbf{k} ^{\prime}\mathbf{k%
} }$ reduces to 
\begin{equation}  \label{eq:D}
\mathit{D} _{\mathbf{k} ^{\prime}\mathbf{k} }=\delta_{\mathbf{k} ^{\prime}%
\mathbf{k} }\left(a^{\dagger}_{\mathbf{k}}a^{}_{\mathbf{k}
}+b^{\dagger}_{-\mathbf{k} }b^{}_{-\mathbf{k}
}\right) 
\end{equation}
This operator which would be  zero for fermion pairs  taken as
elementary bosons, allows us to generate the dimensionless Pauli scatterings for fermion
exchanges between composite bosons in the absence of fermion interaction. Following our works on excitons\cite%
{CobosonPhysicsReports}, these are formally defined through 
\begin{equation}\label{eq:dbeta}
\left[\mathit{D} _{\mathbf{k} ^{\prime}_1\mathbf{k} ^{}_1},\beta^{\dagger}_{%
\mathbf{k} _2}\right]  =\sum_{\mathbf{k} ^{\prime}_2}\left\{\lambda\left(%
\begin{smallmatrix}\vk'_2&\vk_2\\\vk'_1&\vk_1\end{smallmatrix}\right) 
+\left(\mathbf{k} ^{\prime}_1\leftrightarrow\mathbf{k} ^{\prime}_2\right)
\right\} \beta^{\dagger}_{\mathbf{k} ^{\prime}_2}
\end{equation}
By noting that

\begin{equation}  \label{eq:aBeta}
\left[a^{\dagger}_{\mathbf{k} }a^{}_{\mathbf{k} },\beta^{\dagger}_{\mathbf{p}
}\right]  =\delta_{\mathbf{k} \mathbf{p} }\beta^{\dagger}_{\mathbf{p} }=%
\left[b^{\dagger}_{-\mathbf{k} }b^{}_{-\mathbf{k} },\beta^{\dagger}_{\mathbf{%
p} }\right]  
\end{equation}
it is then easy to show that 
\begin{equation}  \label{eq:Dcom}
\left[\mathit{D} _{\mathbf{k} ^{\prime}_1\mathbf{k} ^{}_1},\beta^{\dagger}_{%
\mathbf{k} _2}\right]  =2\beta^{\dagger}_{\mathbf{k} ^{}_2}\delta_{\mathbf{k} _1%
\mathbf{k} _2}\delta_{\mathbf{k} ^{\prime}_1,\mathbf{k} ^{}_2}
\end{equation}
This leads us to identify the Pauli scattering of two zero-momentum free fermion pairs appearing in Eq.(\ref{eq:dbeta}), with a
product of Kronecker symbols 
\begin{equation}  \label{eq:pauliscattering}
\lambda\left(\begin{smallmatrix}\vk'_2&\vk_2\\\vk'_1&\vk_1\end{smallmatrix}%
\right)  =\delta_{\mathbf{k} ^{\prime}_1\mathbf{k} ^{}_1}\delta_{\mathbf{k}
^{\prime}_2\mathbf{k} ^{}_2}\delta_{\mathbf{k} ^{}_1\mathbf{k} ^{}_2}
\end{equation}
Such a simple expression results from the fact that these pairs are made of two free fermions, but also from the fact that they have one degree of freedom only.
Actually, this Pauli scattering is just the one we expect for fermion exchanges between $\left(\mathbf{k} _1,\mathbf{k} _2\right) $ pairs in the absence of fermion interaction,
as visualized by the Shiva diagram of Fig.(1a). Indeed, from this
diagram, it is clear that we must have $\left(\mathbf{k} ^{\prime}_1=\mathbf{%
k} ^{}_1,\mathbf{k} ^{\prime}_2=\mathbf{k} ^{}_2\right) $ and $\left(-\mathbf{k}
^{\prime}_2=-\mathbf{k} ^{}_1,-\mathbf{k} ^{\prime}_1=-\mathbf{k} ^{}_2\right) $ which just gives the delta
factors of Eq.\eqref{eq:pauliscattering}.

\begin{figure}[htb]
\centering
\par
  \subfloat[][]{\includegraphics[width=0.3\textwidth]{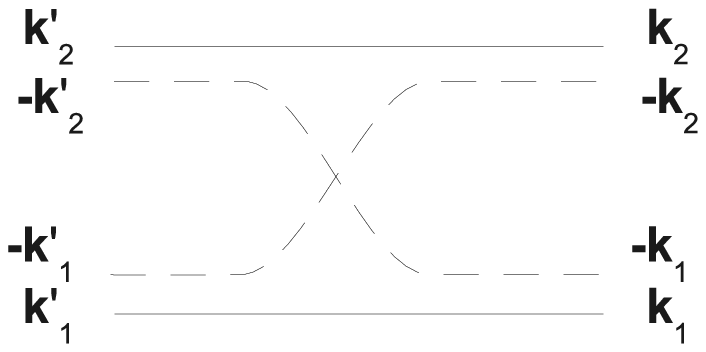}\label{fig:lambda}}\qquad
 \subfloat[][]{\includegraphics[width=0.3\textwidth]{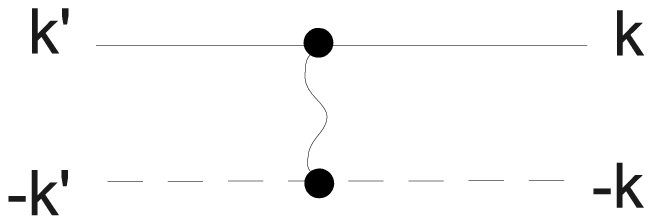}\label{fig:direct}}\\
  \subfloat[][]{\includegraphics[width=0.4\textwidth]{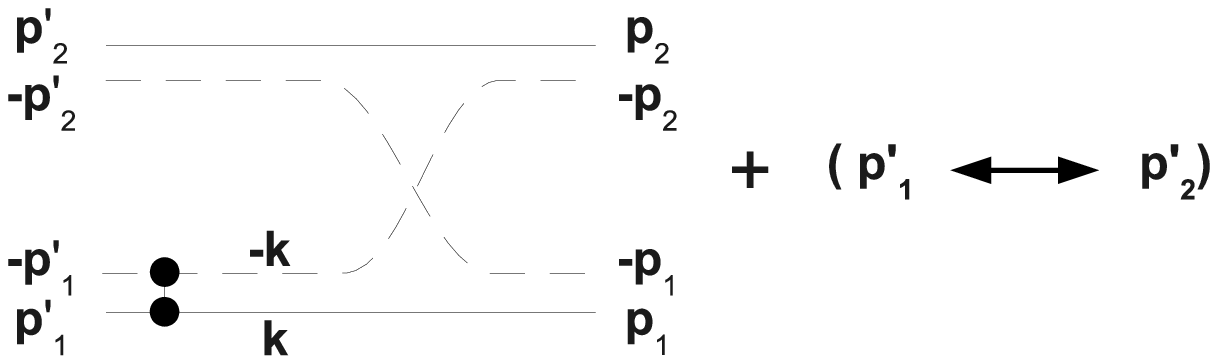}\label{fig:chi}} 
\par
\caption{Shiva diagram of free pairs }
\normalsize
\begin{flushleft}	
\subref{fig:lambda} Pauli scattering $\lambda\left(%
\begin{smallmatrix}\vk'_2&\vk_2\\\vk'_1&\vk_1\end{smallmatrix}\right)  $ for
electron exchange between two free pairs $\left(\mathbf{k} _1,\mathbf{k}
_2\right) $, as given by Eq.\eqref{eq:pauliscattering}. Up spin electrons
are represented by solid lines, down spin electrons by dashed lines. 
\par\subref{fig:direct} The BCS potential given in Eq.\eqref{eq:vbcs}
transforms a $\mathbf{k} $ pair into a $\mathbf{k} ^{\prime}$ pair, with a
constant scattering $-V$, in the case of a separable potential $v_{\mathbf{k}
^{\prime}\mathbf{k} }=-V\,w_{\mathbf{k} ^{\prime}}w_{\mathbf{k} }$.
\par
\subref{fig:chi} Interaction scattering $\chi\left(%
\begin{smallmatrix}\vp'_2&\vp_2\\\vp'_1&\vp_1\end{smallmatrix}\right)  $
between two free pairs, as given in Eq.\eqref{eq:interactSc}. Since the BCS
potential acts within one pair only, scattering between two pairs can
only come from exchange induced by the Pauli exclusion principle. 
\end{flushleft}

\end{figure}

\subsection{Interaction scattering}

We now turn to the interaction scatterings resulting from fermion-fermion interaction. For a free fermion hamiltonian 
\begin{equation}  \label{eq:h0}
H_0=\sum{\epsilon_\vk\left(a^{\dagger}_{\mathbf{k} } a^{}_{\mathbf{k}
}+b^{\dagger}_{\mathbf{k} } b^{}_{\mathbf{k} }\right) }
\end{equation}
Eq.\eqref{eq:aBeta} readily gives 
\begin{equation}  \label{eq:betaH}
\left[H_0,\beta^{\dagger}_\vp\right]  =2\epsilon_\vp\beta^{\dagger}_\vp
\end{equation}

In  standard BCS superconductivity, these fermion pairs interact through the reduced potential
\begin{equation}  \label{eq:vbcs}
V_{BCS}=\sum{v_{\mathbf{k} ^{\prime}\mathbf{k} }\beta^{\dagger}_{\mathbf{k}
^{\prime}}\beta^{}_{\mathbf{k} }}
\end{equation}
We will show below that this potential must be taken as separable $v_{\mathbf{k} ^{\prime}\mathbf{k}}=-Vw_{\mathbf{k} ^{\prime} }w_\vk$ with moreover $w_\vk^2=w_\vk$ in order to possibly find the $N$-pair eigenstates of $H_0+V_{BCS}$ analytically.

It is of importance to note that this potential fundamentally is a (1x1) potential in the fermion pair subspace since fermion $\mathbf{k}$ 
interacts with one fermion only of the other
species, namely fermion $\left(-\mathbf{k} \right)$(see Fig.(1b)). As a crucial consequence, this prevents direct interaction between two zero-momentum pairs. The only way these pairs feel each other, i.e., interact in the most general sense, is through the Pauli exclusion principle.

  For this (1x1)
potential, we do have 
\begin{equation}  \label{eq:vbeta}
\left[V_{BCS},\beta^{\dagger}_\vp\right] 
=\gamma^{\dagger}_\vp+V^{\dagger}_\vp
\end{equation}
where $\gamma^{\dagger}_\vp=\sum_\vk\beta^{\dagger}_\vk{}v_{%
\mathbf{k} \mathbf{p} }$ while $V^{\dagger}_\vp$, that we will call ``creation potential''  of the free fermion pair 
$\mathbf{p} $, is given by 
\begin{equation}  \label{eq:betaV}
V^{\dagger}_\vp=-{\gamma^{\dagger}_\vp}\left(a^{\dagger}_{\mathbf{p} }a^{}_{%
\mathbf{p} }+b^{\dagger}_{-\mathbf{p} }b^{}_{-\mathbf{p} }\right) 
\end{equation}
The general property of creation potentials is that they give zero when acting on vacuum. As now shown, this operator allows us to generate the interactions of the $\vp$ pair with the rest of the system.

While the $\gamma^{\dagger}_\vp$ part of Eq.\eqref{eq:vbeta} 
commutes with $%
\beta^{\dagger}_{\vp ^\prime}$, 
this is not so for the creation potential $%
V^{\dagger}_\vp$. Using Eq. (\ref{eq:aBeta}), its commutator precisely reads 
\begin{equation}\begin{split}  \label{eq:vpotbeta}
\left[V^{\dagger}_{\mathbf{p} _1},\beta^{\dagger}_{\mathbf{p} _2}\right] 
&=-2\delta_{\mathbf{p} _1\mathbf{p} _2}\gamma^{\dagger}_{\mathbf{p}
_1}\beta^{\dagger}_{\mathbf{p} _1}\\
&=-2\delta_{\mathbf{p} _1\mathbf{p} _2}\sum_\mathbf{k}\beta^{\dagger}_{\mathbf{k}}\beta^{\dagger}_{\mathbf{p} _1}v_{\mathbf{k}\mathbf{p} _1}
\end{split}\end{equation}
This allows us to identify the interaction scattering for zero-momentum free pairs,
formally defined as \cite{CobosonPhysicsReports}
\begin{equation}  \label{eq:vBeta}
\left[V^{\dagger}_{\mathbf{p} _1},\beta^{\dagger}_{\mathbf{p} _2}\right] 
=\sum\chi\left(\begin{smallmatrix}\vp'_2&\vp_2\\\vp'_1&\vp_1\end{smallmatrix}%
\right)  \beta^{\dagger}_{\mathbf{p} ^{\prime}_1}\beta^{\dagger}_{\mathbf{p}
^{\prime}_2}
\end{equation}
with a sequence of one (2x2) fermion exchange between two pairs and one (1x1) fermion interaction inside one pair. Indeed, this sequence leads to
\begin{equation}  \label{eq:interactSc}
\begin{split}
\chi\left(\begin{smallmatrix}\vp'_2&\vp_2\\\vp'_1&\vp_1\end{smallmatrix}%
\right)  &=-\sum_\vk\left\{v_{\mathbf{p} ^{\prime}_1\mathbf{k} }\lambda\left(%
\begin{smallmatrix}\vp'_2&\vp_2\\\vk&\vp_1\end{smallmatrix}\right)  +\left(%
\mathbf{p} ^{\prime}_1\leftrightarrow\mathbf{p} ^{\prime}_2\right) \right\} 
\\
&=-\left(v_{\mathbf{p} ^{\prime}_1,\mathbf{p} _1}\delta_{\mathbf{p}
^{\prime}_2,\mathbf{p} _2}+v_{\mathbf{p} ^{\prime}_2,\mathbf{p} _2}\delta_{%
\mathbf{p} ^{\prime}_1,\mathbf{p} _1}\right) \delta_{\mathbf{p} _2,\mathbf{p}
_1}
\end{split}%
\end{equation}
When inserted into Eq. (\ref{eq:vBeta}), this readily gives Eq. (\ref{eq:vpotbeta}). 

This interaction scattering is visualized by the diagram of Fig.(1c): the free pairs $\mathbf{p}_1$ and $\mathbf{p}_2$ first
exchange an electron. As for any exchange, this brings a minus sign. In a
second step, the electrons of one of the two pairs, $\mathbf{p}' _1$ or ${\mathbf{p}}' _2$,  interact via the BCS
potential. It is clear that, since the BCS potential has a (1x1)
structure within the pair subspace, the scattering between two pairs can only result, as ahead said, from
electron exchange between pairs, i.e., Pauli blocking. This diagram evidences it.

It is worth noting that electron exchange and electron interaction do not play a symmetrical role in this interaction scattering. Indeed, process in which the interaction takes place before the exchange - instead of after as in Fig.1c - would lead to
\begin{equation}
-\sum_\vk \lambda\left(%
\begin{smallmatrix}\vp'_2&\vp_2\\\vp'_1&\mathbf{k}\end{smallmatrix}\right)v_{\mathbf{k}\mathbf{p} _1 } =-\delta_{\mathbf{p}
^{\prime}_1,\mathbf{p}^{\prime} _2}\delta_{\mathbf{p}
^{\prime}_2,\mathbf{p} _2}v_{\mathbf{p} _2\mathbf{p} _1}
\end{equation}
which is definitely different from the first term of Eq.(19).

In the next section, we use this commutation formalism to rederive the equations that Richardson has obtained for the eigenstates of $N$ Cooper pairs through a totally different route. The new derivation we have proposed, through its diagrammatic support, enlightens some important physical aspects of this exactly solvable problem.

\section{Richardson's equations for N Cooper pairs\label{sec:rich}}

In order to better grasp how these equations develop, we are going to increase the number of pairs in the potential layer one by one, starting from a single pair.

\subsection{One pair}

We first consider a state in which one free pair $(\mathbf{k} ,-\mathbf{k} )$ is added to a
``frozen'' Fermi sea $\left|F_0\right> $, i.e., a sea which does not feel the BCS potential.
This means that the $v_{\mathbf{k} ^{\prime}\mathbf{k} }$ prefactors in Eq.%
\eqref{eq:vbcs} cancel for all $\mathbf{k} $ belonging to $\left|F_0\right> $ in order to have $V_{BCS} \left|F_0\right>=0$.

 Note that this ``one-pair'' state actually contains $N_0+1$ electron pairs, 
$N_0$ being the number of pairs in the frozen sea, so that this state is a many-body state already, but in the most simple sense since the Fermi sea $%
\left|F_0\right> $ is just there to block states by the Pauli exclusion
principle. This Fermi sea mainly brings a finite density of state for all
states above it, a crucial point to have a bound state in 3D whatever the weakness of the attracting BCS potential.

By choosing the zero energy such
that $H_0\left|F_0\right> =0$, Eqs.(\ref{eq:betaH},\ref{eq:vbeta}) gives the hamiltonian $H=H_0+V_{BCS}$
acting on a one-free-pair state as 
\begin{equation}
H\beta^{\dagger}_\vk\left|F_0\right>  =\left[H,\beta^{\dagger}_\vk\right] 
\left|F_0\right> 
=\left(2\epsilon_\vk\beta^{\dagger}_\vk+\gamma^{\dagger}_\vk+V^{\dagger}_\vk%
\right) \left|F_0\right>  
\end{equation}
We then note that 
\begin{equation}\label{eq:Vk0}
V^{\dagger}_\vk\left|F_0\right> =0
\end{equation}
since the $v_{\mathbf{k} \mathbf{p} }$ factor included
in the $\gamma^{\dagger}_{\mathbf{k} }$ part of $V^{\dagger}_\vk$,  brings
 $v_{\mathbf{k} \mathbf{p} }a^{\dagger}_{\mathbf{p} }a^{}_{%
\mathbf{p} }\left|F_0\right>=v_{\mathbf{k} \mathbf{p} }b^{\dagger}_{-\mathbf{p} }b^{}_{-\mathbf{p} }\left|F_0\right>=0$;
 
Next, we subtract $E _1\beta^{\dagger}_\vk\left|F_0\right>  $ to
the two sides of the above equation, with $E_1$ yet undefined, but assumed to be different from any $2\epsilon_{\mathbf{k}}$.  We then divide the resulting equation by $%
\left(2\epsilon_\vk-E _1\right) $.  This gives
\begin{equation}  \label{eq:HE1}
 (H-E_1)\frac{1}{2\epsilon_\vk-E _1} \beta^{\dagger}_\vk%
\left|F_0\right>  =\beta^{\dagger}_\vk\left|F_0\right>  +\frac{1}{%
2\epsilon_\vk-E _1} \gamma^{\dagger}_\vk\left|F_0\right>  
\end{equation}

To go further and possibly obtain the one-pair eigenstate of the hamiltonian $H$
in a compact analytical form, it is necessary to approximate the BCS potential by a separable potential $v_{\mathbf{k} \mathbf{p} }=-V\,w_\vk{}w_\vp$.
The operator $\gamma^\dagger_\vk$ in Eq. \eqref{eq:vbeta} then reduces to  
\begin{equation}\gamma^{\dagger}_\vk=-V\,w_\vk\beta^{\dagger}
\end{equation} 
where $\beta^\dagger$ is given by
\begin{equation}  \label{eq:gammaBeta}
\beta^{\dagger}=\sum_%
\vp{}w_\vp\beta^{\dagger}_\vp
\end{equation}
If we  now multiply Eq.\eqref{eq:HE1} by $w_\vk$ and sum over $\mathbf{k} $,
we end with 
\begin{equation}\label{eq:1pair}
(H-E _1)B^{\dagger}(E _1)\left|F_0\right>  =\left[1-V\sum_\vk{%
\frac{w_\vk^2}{2\epsilon_\vk-E _1}}\right]
\beta^{\dagger}\left|F_0\right>  
\end{equation}
where the operator $B^{\dagger}(E)$ is defined as  
\begin{equation}  \label{eq:B}
B^{\dagger}(E)=\sum_\vk{B_\vk^{\dagger}(E)}\quad\quad B_\vk^{\dagger}(E)=\frac{w_\vk}{2\epsilon_\vk-E}\beta^{\dagger}_\vk
\end{equation}

Eq.(\ref{eq:1pair}) readily shows that  $B^{\dagger}(%
E _1)\left|F_0\right> $ is  one-pair eigenstate of the hamiltonian $H$ with energy  $%
E _1$, provided that the bracket in the RHS is zero, i.e., $E_1$  fulfills
\begin{equation}  \label{eq:SchOne}
1=V\sum_\vk{\frac{w_\vk^2}{2\epsilon_\vk-E _1}}
\end{equation}
This is just the well-known equation for the single pair energy
derived by Cooper.

\subsection{Two pairs}

We now add two pairs to the frozen sea $\left|F_0\right>$. Eqs.(\ref{eq:betaH},\ref{eq:vbeta}) yield 
\begin{equation}  \label{eq:SchTwo}
\begin{split}
H\beta^{\dagger}_{\mathbf{k} _1}\beta^{\dagger}_{\mathbf{k}
_2}\left|F_0\right>   &=\left(\left[H,\beta^{\dagger}_{\mathbf{k} _1}\right]
\beta^{\dagger}_{\mathbf{k} _2}+\beta^{\dagger}_{\mathbf{k} _1}\left[%
H,\beta^{\dagger}_{\mathbf{k} _2}\right]  \right) \left|F_0\right>   \\
&=\left(2\epsilon_{\mathbf{k} _1}+2\epsilon_{\mathbf{k} _2}\right)
\beta^{\dagger}_{\mathbf{k} _1}\beta^{\dagger}_{\mathbf{k}
_2}\left|F_0\right>\\
&\quad\:   +\left|V_{\mathbf{k} _1\mathbf{k} _2}\right>+\left|W_{\mathbf{k} _1\mathbf{k} _2}\right>  
\end{split}%
\end{equation}
The last two terms come from interactions between the four electrons of the two pairs.  The first term of Eq. (\ref{eq:vbeta}) readily gives $\left|V_{\mathbf{k} _1\mathbf{k} _2}\right> $ as 
\begin{equation}
\begin{split}
\left|V_{\mathbf{k} _1\mathbf{k} _2}\right>& =\left(\gamma^{\dagger}_{\mathbf{%
k} _1}\beta^{\dagger}_{\mathbf{k} _2}+\gamma^{\dagger}_{\mathbf{k}
_2}\beta^{\dagger}_{\mathbf{k} _1}\right) \left|F_0\right> \\
&=-V\left(w_{\mathbf{k} _1}\beta^{\dagger}_{\mathbf{k} _2}+\omega^{\dagger}_{\mathbf{k}
_2}\beta^{\dagger}_{\mathbf{k} _1}\right)\beta^\dagger \left|F_0\right>  
\end{split}
\end{equation}
The second term of Eq. (\ref{eq:vbeta}) yields
\begin{equation}
\left|W_{\mathbf{k} _1\mathbf{k} _2}\right> =\left(V^{\dagger}_{\mathbf{%
k} _1}\beta^{\dagger}_{\mathbf{k} _2}+\beta^{\dagger}_{\mathbf{k}
_1}V^{\dagger}_{\mathbf{k} _2}\right) \left|F_0\right> 
\end{equation}
To calculate it, we again use commutators.  Since  $V^{\dagger}_{\mathbf{k}}$ acting on the frozen sea $\left|F_0\right>$ gives zero (see Eq. (\ref{eq:Vk0})), we find from Eqs. (\ref{eq:vBeta},\ref{eq:interactSc}) 
\begin{equation}\label{eq:omega2}
\begin{split}
\left|W_{\mathbf{k} _1\mathbf{k} _2}\right>&=\sum_{\mathbf{p} ^{\prime}_1\mathbf{p}
^{\prime}_2}\chi\left(\begin{smallmatrix}\vp'_2&\vk_2\\\vp'_1&\vk_1%
\end{smallmatrix}\right)  \beta^{\dagger}_{\mathbf{p} ^{\prime}_1}\beta^{%
\dagger}_{\mathbf{p} ^{\prime}_2}\left|F_0\right> \\
&=2V\delta_{\mathbf{k} _1\mathbf{k} _2}w_{\mathbf{k} _1}\beta^{\dagger}_{
\mathbf{k} _1} \beta^{\dagger}\left|F_0\right> 
\end{split}
\end{equation}

\begin{figure}[htb]
   \includegraphics[width=0.45\textwidth]{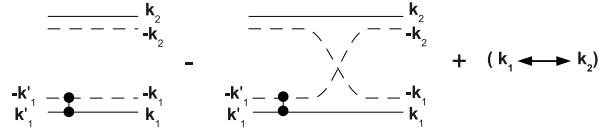}
\caption{Shiva diagram for the interaction part $\left|V_{\mathbf{k} _1\mathbf{k} _2}\right>+\left|W_{\mathbf{k} _1\mathbf{k} _2}\right>$ of the Hamiltonian $H$ acting on two free pairs, as given in Eqs.(30)and (32)}
\label{fig:twoP}
\end{figure}

The interaction part of $H$ acting on two free pairs is  visualized by
the diagram of Fig. \ref{fig:twoP}. In $\left|V_{\mathbf{k} _1\mathbf{k} _2}\right>$, one pair stays unchanged while the other pair suffers a BCS interaction. In $\left|W_{\mathbf{k} _1\mathbf{k} _2}\right>$, the two pairs exchange an electron and then one pair interacts.   These diagrams evidence the fact that,
due to the (1x1) structure of the BCS potential,  two pairs 
can  interact by fermion exchange only as a result of the
Pauli exclusion principle.

To go further, we subtract $E _2\beta^{\dagger}_{\mathbf{k}
_1}\beta^{\dagger}_{\mathbf{k} _2}\left|F_0\right>  $ to the two sides of Eq.%
\eqref{eq:SchTwo}, with $E _2$ yet undefined. We split $E _2$ as $R_1+R_2$ and we multiply
the resulting equation by \\$w_{\mathbf{k} _1}w_{\mathbf{k} _2}/\left(2%
\epsilon_{\mathbf{k} _1}-R_1\right) \left(2\epsilon_{\mathbf{k}
_2}-R_2\right) $. This yields

\begin{multline}  \label{eq:SchTwo2}
(H-E _2)B^{\dagger}_{\mathbf{k} _1}(R_1)B^{\dagger}_{\mathbf{k}
_2}(R_2)\left|F_0\right>   = \\
\left\{B^{\dagger}_{\mathbf{k} _1}(R_1)\left[w_{\mathbf{k}
_2}\beta^{\dagger}_{\mathbf{k} _2}-\frac{Vw_{\mathbf{k} _2}^2}{2\epsilon_{%
\mathbf{k} _2}-R_2}\beta^{\dagger}\right]+(1\leftrightarrow2)\right\}
\left|F_0\right>  \\
+2V\left[\delta_{{\mathbf{k} _1}{\mathbf{k} _2}}\frac{w^3_{{\mathbf{k} _1}}}{\left(2%
\epsilon_{\mathbf{k} _1}-R_1\right) \left(2\epsilon_{\mathbf{k}
_1}-R_2\right)}\beta^\dagger_{{\mathbf{k} _1}}\right]\beta^\dagger\left|F_0\right>  
\end{multline}
the last term coming from the exchange interaction term $\left|W_{\mathbf{k} _1\mathbf{k} _2}\right>$

As a last step, we sum over $(\vk_1,\vk_2)$.  The sum over $(\vk_2)$ in the first bracket readily gives 
\begin{equation}\label{eq:1-v}
\left(1-V\sum\frac{w_{\mathbf{k} }^2}{2\epsilon_{\mathbf{k} }-R_2}\right)\beta^\dagger
\end{equation}
To calculate the sum over $(\vk_1,\vk_2)$ in the second bracket, we first note that
\begin{equation}\label{eq:inverse}
\begin{split}
&\frac{1}{\left(2\epsilon_{\mathbf{k} _1}-R_1\right)
\left(2\epsilon_{\mathbf{k} _1}-R_2\right)}\\&=\left[
\frac{1}{\left(2\epsilon_{\mathbf{k} _1}-R_1\right)}-\frac{1}{\left(2\epsilon_{\mathbf{k}
_1}-R_2\right) }\right]\frac1{\left(R_1-R_2\right) } 
\end{split}
\end{equation}
which is valid, provided that $R_1\neq{}R_2$,
a condition that we can always enforce since the unique requirement is to have $R_1+R_2=E_2$. For $w^2_{{\mathbf{k} }}=w_{{\mathbf{k} }}$, we then find 
\begin{multline}\label{eq:deltakk}
\sum_{{\mathbf{k} _1}{\mathbf{k} _2}}\delta_{{\mathbf{k} _1}{\mathbf{k} _2}}\frac{w^3_{{\mathbf{k} _1}}}{\left(2%
\epsilon_{\mathbf{k} _1}-R_1\right) \left(2\epsilon_{\mathbf{k}
_1}-R_2\right)}\beta^\dagger_{{\mathbf{k} _1}}\\=\frac1{\left(R_1-R_2\right) }[B^\dagger(R_1)-B^\dagger(R_2)]
\end{multline}
 Summation over $(\vk_1,\vk_2)$ of Eq. \eqref{eq:SchTwo2} then yields 
\begin{multline}  \label{eq:SchTwo3}
(H-E_2)B^{\dagger}(R_1)B^{\dagger}(R_2)\left|F_0\right>   = \\
\left\{B^{\dagger}(R_1)\left[1-V\sum\frac{w_{\mathbf{k} }^2}{2\epsilon_{%
\mathbf{k} }-R_2}+\frac{2V}{R_1-R_2}\right] +(1\leftrightarrow2)\right\}  \\
\beta^{\dagger}\left|F_0\right>  
\end{multline}

The above equation evidences that $B^{\dagger}(R_1)B^{\dagger}(R_2)%
\left|F_0\right>  $ is two-pair eigenstate of the hamiltonian $H$ with  energy $%
E _2=R_1+R_2$ provided that the bracket in the above equations is zero, i.e., $\left(R_1,R_2\right) $ fulfill two
equations, known as Richardson's equations for two pairs 
\begin{equation}
1=V\sum\frac{w_{\mathbf{k} }^2}{2\epsilon_{\mathbf{k} }-R_1}+\frac{2V}{R_1-R_2}%
=(1\leftrightarrow2)
\end{equation}

\subsection{Three pairs}

We now turn to three pairs in order to see how these equations develop for an
increasing number of pairs. Two usually is not generic, while three most often is.  We will here see that when $H$ acts on three pairs, one at least among the three pairs stays unchanged.   This is a step toward understanding why the $N$ dependence of the $N$-pair ground state energy is in $N(N-1)$ only, with no term in $N(N-1)(N-2)$ and higher, as the validity of our low density result extrapolated to the high density BCS regime, seems to indicate.

We start with 
\begin{eqnarray}  \label{eq:SchThree}
&&H\beta^{\dagger}_{\mathbf{k} _1}\beta^{\dagger}_{\mathbf{k}
_2}\beta^{\dagger}_{\mathbf{k} _3}\left|F_0\right>  \hspace{5cm}
\nonumber\\
&=&\left\{\left[H,\beta^{\dagger}_{\mathbf{k} _1}\right]  \beta^{\dagger}_{%
\mathbf{k} _2}\beta^{\dagger}_{\mathbf{k} _3}+\beta^{\dagger}_{\mathbf{k} _1}%
\left[H,\beta^{\dagger}_{\mathbf{k} _2}\right]  \beta^{\dagger}_{\mathbf{k}
_3}\right.
\nonumber\\ &&\left.
+\beta^{\dagger}_{\mathbf{k} _1}\beta^{\dagger}_{\mathbf{k} _2}\left[%
H,\beta^{\dagger}_{\mathbf{k} _3}\right]  \right\}
\left|F_0\right> 
\end{eqnarray}%
 Using Eqs.(\ref{eq:betaH},\ref{eq:vbeta}), we again  split the above equation into a kinetic part and two interaction parts
\begin{equation}  \label{eq:SchThree2}
\begin{split}
H\beta^{\dagger}_{\mathbf{k} _1}\beta^{\dagger}_{\mathbf{k}
_2}\beta^{\dagger}_{\mathbf{k} _3}\left|F_0\right>   &=\left(2\epsilon_{%
\mathbf{k} _1}+2\epsilon_{\mathbf{k} _2}+2\epsilon_{\mathbf{k} _3}\right)
\beta^{\dagger}_{\mathbf{k} _1}\beta^{\dagger}_{\mathbf{k}
_2}\beta^{\dagger}_{\mathbf{k} _3}\left|F_0\right>   \\
&+\left|{V}_{\mathbf{k} _1\mathbf{k} _2\mathbf{k} _3}\right> +\left|{W}_{\mathbf{k} _1\mathbf{k} _2\mathbf{k} _3}\right> 
\end{split}%
\end{equation}
As in the case of two pairs, the  BCS potential generates direct processes which are given by 
\begin{equation}  \label{eq:vThree}
\left|V_{\mathbf{k} _1\mathbf{k} _2\mathbf{k} _3}\right> =
\left(\gamma^{\dagger}_{\mathbf{k} _1}\beta^{\dagger}_{\mathbf{k}
_2}\beta^{\dagger}_{\mathbf{k} _3}+\gamma^{\dagger}_{\mathbf{k}
_2}\beta^{\dagger}_{\mathbf{k} _3}\beta^{\dagger}_{\mathbf{k}
_1}+\gamma^{\dagger}_{\mathbf{k} _3}\beta^{\dagger}_{\mathbf{k}
_1}\beta^{\dagger}_{\mathbf{k} _2}\right) \left|F_0\right>   
\end{equation}
since $\gamma^{\dagger}_{\mathbf{k}}$ and $\beta^{\dagger}_{\mathbf{k}'}$ commute.  It also generates exchange processes, which appear as 
\begin{equation}
\left|W_{\mathbf{k} _1\mathbf{k} _2\mathbf{k} _3}\right> =\left(V^{\dagger}_{\mathbf{k} _1}\beta^{\dagger}_{\mathbf{k}
_2}\beta^{\dagger}_{\mathbf{k} _3}+\beta^{\dagger}_{\mathbf{k}
_1}V^{\dagger}_{\mathbf{k} _2}\beta^{\dagger}_{\mathbf{k} _3}+\beta^{%
\dagger}_{\mathbf{k} _1}\beta^{\dagger}_{\mathbf{k} _2}V^{\dagger}_{\mathbf{k%
} _3}\right) \left|F_0\right>  
\end{equation}
To calculate them, we again use commutators and the fact that $V^{\dagger}_\vk\left|F_0\right>  =0$. 
Eq. \eqref{eq:vBeta} allows us to rewrite  $\left|W_{\mathbf{k} _1\mathbf{k} _2\mathbf{k} _3}\right>$ in a more symmetrical form as
\begin{equation}  \label{eq:vThree2}
\begin{split}
&\left\{\left[V^{\dagger}_{\mathbf{k} _1},\beta^{\dagger}_{\mathbf{k} _2}%
\right]  \beta^{\dagger}_{\mathbf{k} _3}+\beta^{\dagger}_{\mathbf{k} _2}%
\left[V^{\dagger}_{\mathbf{k} _1},\beta^{\dagger}_{\mathbf{k} _3}\right] 
+\beta^{\dagger}_{\mathbf{k} _1}\left[V^{\dagger}_{\mathbf{k}
_2},\beta^{\dagger}_{\mathbf{k} _3}\right]  \right\} \left|F_0\right>   \\
=&\sum_{\vk^{\prime}_1\mathbf{k} ^{\prime}_2}\beta^{\dagger}_{\mathbf{k}
^{\prime}_1}\beta^{\dagger}_{\mathbf{k} ^{\prime}_2} \\
&\left\{\chi\left(\begin{smallmatrix}\vk'_2&\vk_2\\\vk'_1&\vk_1%
\end{smallmatrix}\right)  \beta^{\dagger}_{\mathbf{k} _3}+\chi\left(%
\begin{smallmatrix}\vk'_2&\vk_3\\\vk'_1&\vk_2\end{smallmatrix}\right) 
\beta^{\dagger}_{\mathbf{k} _1}+\chi\left(\begin{smallmatrix}\vk'_2&\vk_1\
\\\vk'_1&\vk_3\end{smallmatrix}\right)  \beta^{\dagger}_{\mathbf{k}
_2}\right\} \left|F_0\right>  
\end{split}%
\end{equation}

\begin{figure}[htb]
   \includegraphics[width=0.4\textwidth]{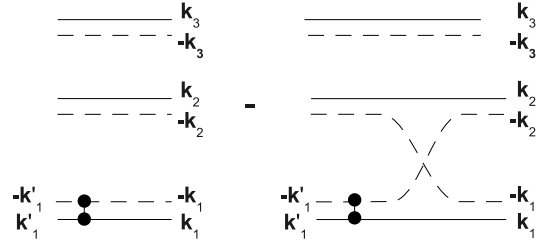}
\caption{Shiva diagram for the interaction part $\left|V_{\mathbf{k} _1\mathbf{k} _2
\mathbf{k} _3}\right> +\left|W_{\mathbf{k} _1\mathbf{k} _2
\mathbf{k} _3}\right>$ of the Hamiltonian $H$ acting on three pairs }\label{fig:threeP}
 \end{figure}

The interaction part of Eq. (\ref{eq:SchThree2}), namely $\left|V_{\mathbf{k} _1\mathbf{k} _2%
\mathbf{k} _3}\right>+\left|W_{\mathbf{k} _1\mathbf{k} _2\mathbf{k} _3}\right> $, is represented by the diagrams of Fig.\ref{fig:threeP}. \\$\left|V_{\mathbf{k} _1\mathbf{k} _2
\mathbf{k} _3}\right>$ has interactions inside a single pair, two pairs
staying unchanged. $\left|W_{\mathbf{k} _1\mathbf{k} _2\mathbf{k} _3}\right> $ contains processes in which the pair suffering the potential exchanges one of its electrons with a second pair, the third pair staying unchanged.  There  are three similar  contributions, obtained by circular permutations.

Using Eq.\eqref{eq:interactSc} for the interaction scattering, we end with

\begin{equation}\label{eq:omega3}
\left|W_{\mathbf{k} _1\mathbf{k} _2\mathbf{k} _3}\right>= 2V(\delta_{\mathbf{k} _1\mathbf{k} _2}w_{\mathbf{k} _1}\beta^{\dagger}_{\mathbf{k} _1}\beta^{\dagger}_{\mathbf{k} _3}+ \text{2 perm.})\beta^{\dagger}\left|F_0\right>
\end{equation}
which has close similarity with  $\left|W_{\mathbf{k} _1\mathbf{k} _2}\right> $ given in Eq. (\ref{eq:omega2}).

To go further, we insert Eq.(\ref{eq:vThree}) for $\left|V_{\mathbf{k} _1\mathbf{k} _2\mathbf{k} _3}\right>$  and Eq.(\ref{eq:omega3}) for $\left|W_{\mathbf{k} _1\mathbf{k} _2\mathbf{k} _3}\right> $ into \eqref{eq:SchThree2}; we subtract $E
_3\beta^{\dagger}_{\mathbf{k} _1}\beta^{\dagger}_{\mathbf{k}
_2}\beta^{\dagger}_{\mathbf{k} _3}\left|F_0\right>  $ to both sides, with $%
E _3$ written as $R_1+R_2+R_3$, and we multiply the resulting equation
by \\$w_{\mathbf{k} _1}w_{\mathbf{k} _2}w_{\mathbf{k} _3}/\left(2\epsilon_{%
\mathbf{k} _1}-R_1\right) \left(2\epsilon_{\mathbf{k} _2}-R_2\right)
\left(2\epsilon_{\mathbf{k} _3}-R_3\right) $. This \\yields

\begin{eqnarray}  \label{eq:SchThree3}
&&(H-E _3)B^{\dagger}_{\mathbf{k} _1}(R_1)B^{\dagger}_{\mathbf{k}
_2}(R_2)B^{\dagger}_{\mathbf{k} _3}(R_3)\left|F_0\right>  \nonumber\\
&=& \Bigl\{B^{\dagger}_{\mathbf{k} _1}(R_1)B^{\dagger}_{\mathbf{k}
_2}(R_2)\left[w_{\mathbf{k} _3}\beta^{\dagger}_{\mathbf{k} _3}-\frac{Vw_{%
\mathbf{k} _3}^2}{2\epsilon_{\mathbf{k} _2}-R_3}\beta^{\dagger}\right] \nonumber\\
&&+\text{2 perm.} \Bigr\}\left|F_0\right> \nonumber\\
&&+2V\Bigl\{B^{\dagger}_{\mathbf{k} _3}(R_3)\left[\frac{\delta_{\mathbf{k} _1\mathbf{%
k} _2}w^3_{\mathbf{k} _1}}{\left(2\epsilon_{\mathbf{k} _1}-R_1\right)
\left(2\epsilon_{\mathbf{k} _1}-R_2\right) }\beta^{\dagger}_{\mathbf{k} _1}\right]\nonumber\\
&&\qquad+\text{2 perm.}\Bigr\} 
\beta^{\dagger}\left|F_0\right> 
\end{eqnarray}
We then sum over $\left(\mathbf{k} _1,\mathbf{k} _2,\mathbf{k} _3\right) $. By calculating the sums of the two brackets as for two pairs, Eqs. (\ref{eq:1-v}) and (\ref{eq:deltakk}) then yield
\begin{multline}  \label{eq:SchThree4}
(H-E _3)B^{\dagger}(R_1)B^{\dagger}(R_2)B^{\dagger}(R_3)\left|F_0%
\right>  = \\
\{B^{\dagger}(R_2)B^{\dagger}(R_3) \\
\left[1-V\sum\frac{w_{\mathbf{k} _1}^2}{2\epsilon_{\mathbf{k} _1}-R_1}-\frac{2V%
}{R_1-R_2}+\frac{2V}{R_3-R_1}\right]  \\
+\text{2 perm.}\}\beta^{\dagger}\left|F_0\right>  
\end{multline}

This leads us to  conclude that the three-pair state $%
B^{\dagger}(R_1)B^{\dagger}(R_2)B^{\dagger}(R_3)\left|F_0\right>  $ is
eigenstate of the hamiltonian $H$ with the energy $E _3=R_1+R_2+R_3$,
provided that $\left(R_1,R_2, R_3\right) $ fulfill the three equations, 
\begin{equation}
\begin{split}
1&=V\sum\frac{w_{\mathbf{k} }^2}{2\epsilon_{\mathbf{k} }-R_1}+\frac{2V}{R_1-R_2%
}+\frac{2V}{R_1-R_3} \\
1&=V\sum\frac{w_{\mathbf{k} }^2}{2\epsilon_{\mathbf{k} }-R_2}+\frac{2V}{R_2-R_3%
}+\frac{2V}{R_2-R_1} \\
1&=V\sum\frac{w_{\mathbf{k} }^2}{2\epsilon_{\mathbf{k} }-R_3}+\frac{2V}{R_3-R_1%
}+\frac{2V}{R_3-R_2}
\end{split}%
\end{equation}

\subsection{$N$ pairs}

The above commutation technique can be easily extended to $N$ pairs. As nicely
visualized by the diagrams of Figs.\ref{fig:twoP} and \ref{fig:threeP}, the
effect of the BCS potential on these $N$ pairs splits into direct and exchange processes:
In the direct set, one pair is affected by the (1x1) scattering while the other $N-1
$ pairs stay unchanged. In the exchange set, this pair, before interaction, also 
exchanges one of its electrons with another pair, the remaining $N-2$
pairs staying unchanged. This understanding shows that an increase of pair number above two, does not really change the structure of the equations
since $N-2$ pairs stay unchanged, the pair which exchanges its fermions with the
pair suffering the interaction being just one among $(N-1)$ pairs.

The procedure is rather straightforward once we have understood
that either $(N-1)$ or $(N-2)$ pairs stay unaffected in the BCS interaction process. The
general form of the $N$-pair eigenstates ultimately appears as 
\begin{equation}  \label{eq:SchThreeN}
(H-E _N)B^{\dagger}(R_1)\cdots{}B^{\dagger}(R_N)\left|F_0\right>  =0
\end{equation}
with $E _N=R_1+\cdots+R_N$, these $R_N$'s being solutions of $N$ coupled
equations 
\begin{equation}
1=V\sum\frac{w_{\mathbf{k} }^2}{2\epsilon_{\mathbf{k} }-R_i}+\sum_{j\neq{i}}%
\frac{2V}{R_i-R_j}\quad\qquad \text{for}\; i=\left(1,...,N\right) 
\end{equation}
Let us explicitly derive these $N$ equations following the procedure we have used for three pairs. 

The hamiltonian acting on $N$ pairs can be written in terms of commutators with any of these $N$ pairs as 
\begin{equation}\label{eq:HN}
\begin{split}
&H\beta^{\dagger}_{\mathbf{k} _1}\cdots\beta^{\dagger}_{\mathbf{k}
_N}\left|F_0\right> \\
&=\left[H,\beta^{\dagger}_{\mathbf{k} _1}\right]  \beta^{\dagger}_{%
\mathbf{k} _2}\cdots\beta^{\dagger}_{\mathbf{k} _N}\left|F_0\right> \\
&\quad+\cdots+\beta^{\dagger}_{\mathbf{k} _1}\cdots\beta^{\dagger}_{\mathbf{k} _{i-1}}
\left[H,\beta^{\dagger}_{\mathbf{k} _i}\right]  \beta^{\dagger}_{\mathbf{k}
_{i+1}}\cdots \beta^{\dagger}_{\mathbf{k}_{N}}\left|F_0\right>\\
&\quad+\cdots+\beta^{\dagger}_{\mathbf{k} _1}\cdots\beta^{\dagger}_{\mathbf{k} _{N-1}}
\left[H,\beta^{\dagger}_{\mathbf{k} _N}\right] \left|F_0\right>
\end{split}
\end{equation}
We then use Eqs. (\ref{eq:betaH},\ref{eq:vbeta}) to replace $\left[H,\beta^{\dagger}_{\mathbf{k} _i}\right]$ by $2\epsilon_{\vk_i}\beta^{\dagger}_{\vk_i}+\gamma^{\dagger}_{\vk_i}+V^{\dagger}_{\vk_i}$.  The first two contributions commute with the other $\beta^\dagger_\vk$'s, so that, when inserted into the above equation, they yield
\begin{equation}
2\left(\epsilon_{\mathbf{k} _1}+\cdots+\epsilon_{\mathbf{k} _N}\right)
\prod^N_{i=1}\beta^{\dagger}_{\mathbf{k} _i}\left|F_0\right>+\left|{V}_{\mathbf{k} _1\cdots\mathbf{k} _N}\right> 
\end{equation}
where the direct interaction part is given by 
\begin{equation}  
\left|V_{\mathbf{k} _1\cdots\mathbf{k} _N}\right> =\sum^N_{i=1}\gamma^{\dagger}_{\mathbf{k}_i}\prod_{m\neq{i}}\beta^{\dagger}_{\mathbf{k} _m}\left|F_0\right>   
\end{equation}
which is similar to Eq. (\ref{eq:vThree}). 

 The part with the creation potential $V^\dagger_{\mathbf{k}_i}$ is more cumbersome.  We again calculate it through commutators.  Let us consider one term. We start as
\begin{equation}
\begin{split}
&V^{\dagger}_{\mathbf{k} _{i}}\beta^{\dagger}_{\mathbf{k} _{i+1}}\cdots\beta^{\dagger}_{\mathbf{k}
_N}\left|F_0\right> \\
&=\left[V^{\dagger}_{\mathbf{k} _{i}},\beta^{\dagger}_{\mathbf{k} _{i+1}}\right]  \beta^{\dagger}_{%
\mathbf{k} _{i+2}}\cdots\beta^{\dagger}_{\mathbf{k} _N}\left|F_0\right> \\
&\quad+\cdots+\beta^{\dagger}_{\mathbf{k} _{i+1}}\cdots\beta^{\dagger}_{\mathbf{k} _{j-1}}
\left[V^{\dagger}_{\mathbf{k} _{i}},\beta^{\dagger}_{\mathbf{k} _j}\right]  \beta^{\dagger}_{\mathbf{k}
_{j+1}}\cdots \beta^{\dagger}_{\mathbf{k}_{N}}\left|F_0\right>\\
&\quad+\cdots+\beta^{\dagger}_{\mathbf{k} _{i+1}}\cdots\beta^{\dagger}_{\mathbf{k} _{N-1}}
\left[V^{\dagger}_{\mathbf{k} _{i}},\beta^{\dagger}_{\mathbf{k} _N}\right] \left|F_0\right>
\end{split}
\end{equation}
 $\left[V^{\dagger}_{\mathbf{k} _{i}},\beta^{\dagger}_{\mathbf{k} _j}\right]$ makes appear the  exchange interaction scattering between the $\vk_i$ and $\vk_j$ pairs, so that the above term generate the exchange interaction scatterings between the $\mathbf{k} _{i}$ pair and all the $\mathbf{k} _{j}$ pairs with  $i<j\leq{N}$.  When inserted into Eq. (\ref{eq:HN}), we find all possible exchange interactions between fermion pairs, so that the set of terms with creation potential ultimately gives
\begin{equation}  
\left|W_{\mathbf{k} _1\cdots\mathbf{k} _N}\right>=\sum_{\vp^{\prime}_1\mathbf{p} ^{\prime}_2}\beta^{\dagger}_{\mathbf{p}
^{\prime}_1}\beta^{\dagger}_{\mathbf{p} ^{\prime}_2} \sum_{i<j}\chi\left(\begin{smallmatrix}\vp'_2&\vk_j\\\vp'_1&\vk_i%
\end{smallmatrix}\right)  \prod_{m\neq(i,j)}\beta^{\dagger}_{\mathbf{k} _m} \left|F_0\right>  
\end{equation}
 which is similar to Eq. (\ref{eq:vThree2}).  If we now use Eq. (\ref{eq:omega2}) for the sum over $(\vp^{\prime}_1,\mathbf{p} ^{\prime}_2)$, we end with 
 
  \begin{equation}  
\left|W_{\mathbf{k} _1\cdots\mathbf{k} _N}\right>=2V\beta^\dagger[ \delta_{\vk_1\vk_2}w_{\vk_2}\beta^{\dagger}_{\mathbf{k} _2}\cdots\beta^{\dagger}_{\mathbf{k} _N}+\cdots ]\left|F_0\right>  
\end{equation}
which also reads in a compact form as 
 \begin{equation}  
\left|W_{\mathbf{k} _1\cdots\mathbf{k} _N}\right>=2V\beta^\dagger\left[ \sum_{i<j}\delta_{\vk_i\vk_j}w_{\vk_j}\beta^\dagger_{\vk_j}\prod_{m\neq(i,j)}\beta^{\dagger}_{\mathbf{k} _m} \right]\left|F_0\right>  
\end{equation}

All this leads for $E_N$ written as $R_1+...+R_N$ to

\begin{equation}\label{eq:HN2}
\begin{split}
&(H-E_N)\beta^{\dagger}_{\mathbf{k} _1}\cdots\beta^{\dagger}_{\mathbf{k}
_N}\left|F_0\right> \\
&=[(2\epsilon_{\vk_1}-R_1)+\cdots+(2\epsilon_{\vk_2}-R_N)]\beta^\dagger_{\vk_1}\cdots\beta^\dagger_{\vk_N}\left|F_0\right>  \\
&\quad-V\beta^\dagger \sum_{i}w_{\vk_i}\beta^\dagger_{\vk_j}\prod_{m\neq{}i}\beta^{\dagger}_{\mathbf{k} _m} \left|F_0\right>  \\
&\quad+2V\beta^\dagger\sum_{i<j}\delta_{\vk_i\vk_j}w_{\vk_j}\beta^\dagger_{\vk_j}\prod_{m\neq(i,j)}\beta^{\dagger}_{\mathbf{k} _m} \left|F_0\right>  
\end{split}
\end{equation}
To go further, we do as before: we multiply both sides of the equation by $w_{\vk_1}\cdots{}w_{\vk_N}/(2\epsilon_{\vk_1}-R_1)\cdots(2\epsilon_{\vk_N}-R_N)$ and we sum over $(\vk_1,\cdots,\vk_N)$.  
The LHS readily gives
\begin{equation}
(H-E _N)B^{\dagger} (R_1)\cdots{}B^{\dagger}(R_N)\left|F_0\right> 
\end{equation}
The first term in the RHS, which comes from the free pair kinetic energy, yields
\begin{equation}
\begin{split}
&[(\sum{}w_{\vk_1}\beta^\dagger_{\vk_1})B^\dagger(R_2)\cdots{}B^\dagger(R_N)+\cdots]\left|F_0\right>\\
&=\beta^\dagger\sum_{i=1}^N\prod_{m\neq{i}}B^\dagger(R_m)\left|F_0\right> 
\end{split}
\end{equation}
The first interaction term, induced by direct processes within one pair, readily leads to 
\begin{equation}
\begin{split}
\beta^\dagger\sum_{i=1}^N(-V\sum_{\vk_i}\frac{w_{\mathbf{k} _i}^2}{2\epsilon_{\mathbf{k} _i}-R_i})\prod_{m\neq{i}}B^\dagger(R_m)\left|F_0\right> 
\end{split}
\end{equation}
while contributions coming from exchange interaction processes appear as
\begin{equation}
\begin{split}
2V\beta^\dagger\left[\sum_{\vk_1\vk_2}\delta_{\vk_1\vk_2}\frac{w^3_{\mathbf{k} _1}}{(2\epsilon_{\mathbf{k} _1}-R_1)(2\epsilon_{\mathbf{k} _2}-R_2)}\beta^\dagger_{\vk_1}\right]\\B^\dagger(R_3)\cdots{}B^\dagger(R_N)\left|F_0\right> 
\end{split}
\end{equation}
By using  Eq. (\ref{eq:deltakk}) for the above bracket, we end with  
\begin{multline}  
(H-E _N)B^{\dagger}(R_1)\cdots{}B^{\dagger}(R_N)\left|F_0%
\right>  = \\
\beta^\dagger\sum_{i=1}^N\left[1-V\sum_\vk\frac{w_{\mathbf{k} }^2}{2\epsilon_{\mathbf{k}}-R_i}-\sum_{j\neq{}i}\frac{2V%
}{R_i-R_j}\right]\\
\sum_{m\neq{i}}B^{\dagger}(R_m)\left|F_0\right>  
\end{multline}
This evidences that $B^{\dagger}(R_1)\cdots{}B^{\dagger}(R_N)\left|F_0\right> $ is $N$-pair eigenstate of the hamiltonian $H$, with energy $E_N=R_1+\cdots+R_N$ provided that all the brackets in the above equation cancel. These just are the $N$ Richardson's equations written in Eq.(49).
\section{Physical understanding}

\subsection{Richardson equations and the Pauli exclusion principle}

The above  derivation of Richardson's equations makes crystal clear the parts in these equations which are directly linked
to the Pauli exclusion principle between fermion pairs through electron exchanges. 

From a mathematical point of view, the link is rather obvious: In the
absence of terms in $V/(R_i-R_j)$, the $N$ equations for $R_i$ would reduce to
the same equation \eqref{eq:SchOne}, so that the solution would be $R^{(0)}_i=%
E _1$ for all $i$. The fact that the energy of $N$ pairs differs
from $N$ times the single pair energy $E_1$ entirely comes from the set of $(R_i-R_j)$'s
different from zero.

Physically, the fact that $E _N$ differs from $NE _1$
comes from interactions between Cooper pairs. Due to the (1x1) structure of the BCS
potential within the pair subspace, interaction between pairs can only be mediated by fermion
exchanges as evidenced from the diagram of Fig. (\ref{fig:chi}). 
Consequently, interactions between pairs
are solely the result of the Pauli exclusion principle. This
Pauli blocking mathematically appears through the various $\delta_{\mathbf{p}
^{\prime}\mathbf{p} }$ factors in the Pauli scatterings $\lambda(%
\begin{smallmatrix}\vp^\prime_2&\vp_2\\\vp_1'&\vp_1\end{smallmatrix})  $. These $\delta$ factors are the ones of the $\left|W_{\mathbf{k} _1\cdots\mathbf{k} _N}\right>$ term.  They ultimately lead to the various $(R_i-R_j)$ differences
 in the Richardson's equations, as easy to follow from our procedure.

In short, the Kronecker symbols in the Pauli scatterings of fermion pairs
take care of states which are excluded by the Pauli exclusion principle. They induce the $2V/(R_i-R_j)$ terms
of the Richardson's equations which ultimately makes the energy of $N$ pairs different
from the energy of $N$ independent pairs.

\subsection{Excitons versus Cooper pairs}

An important feature of the ground state energy $E _N$ for $N$ pairs that
this  derivation possibly explains, is the fact that the
part of $E_N$ coming from interaction, namely $E _N-N%
E _1$ depends on $N$ as $N(N-1)$ only, with no higher order dependence. Indeed, Eq.(1)  also reads
\begin{equation}  \label{eq:en}
E _N=NE _1+N(N-1)\left(\frac{1}{\rho_0} +\frac{\epsilon_c}{%
N_\Omega}\right) 
\end{equation}

In order to have terms in the energy in $N(N-1)(N-2)$, we need topologically connected
diagrams between 3 pairs. The diagram of Fig.\ref{fig:threeP} shows that, when $H$ acts on three pairs, one out of them do not participate to the scattering, so that the three pairs are not connected. In the case of $N$ pairs, this is $(N-2)$ out of the $N$ pairs which are not connected. Connections thus seem to exist between two pairs only. 

This however is not enough to explain that higher order terms do not exist in the energy because the energy of $N$ Wannier excitons has terms in $N(N-1)(N-2)$ and higher\cite{monicOdil}  while the hamiltonian acting on $N$ Wannier excitons  also leaves $N-2$ excitons unchanged. Indeed, let $B_i^\dagger$ be the creation operator of the $i$ exciton with energy $E_i$, i.e., $(H-E_i)B^\dagger_i\left|0\right>=0$.  We do have, as in Eq. (\ref{eq:SchThree}), 

\begin{equation}\label{eq:HB3}
\begin{split}
&HB^{\dagger}_{i_1}B^{\dagger}_{i_2}B^{\dagger}_{i_3}\left|0\right>=\\
&\left\{\left[H,B^{\dagger}_{i_1}\right] B^{\dagger}_{i _2}B^{\dagger}_{i_3}+B^{\dagger}_{i_1}%
\left[H,B^{\dagger}_{i_2}\right] B^{\dagger}_{i_3}\right.\\ 
&\left.+B^{\dagger}_{i_1}B^{\dagger}_{i_2}\left[H,B^{\dagger}_{i_3}\right]  \right\}\left|0\right> 
\end{split}
\end{equation} 
To calculate it, we introduce the $i$ exciton creation operator $V^\dagger_i$ defined as
\begin{equation}
\left[H,B^{\dagger}_{i}\right]=E_iB^\dagger_i+V^\dagger_i
\end{equation}
 which has similarity with Eqs. (\ref{eq:betaH},\ref{eq:vbeta}). This operator is such that $V^{\dagger}_{i}\left|0\right>=0$, as readily seen from the above equation acting on vacuum.  From it, we construct the interaction scatterings of two excitons through
 \begin{equation}  
\left[V^{\dagger}_{i},B^{\dagger}_{j}\right] 
=\sum\xi\left(\begin{smallmatrix}n&j\\m&i\end{smallmatrix}%
\right)  B^{\dagger}_{m}B^{\dagger}_{n}
\end{equation}
which is similar to Eq.(\ref{eq:vBeta}). When used into Eq.(\ref{eq:HB3}), this yields
\begin{equation}\label{eq:HEEE}
\begin{split}
&(H-E_{i_1}-E_{i_2}-E_{i_3})B^{\dagger}_{i_1}B^{\dagger}_{i_2}B^{\dagger}_{i_3}\left|0\right>\\
=&\sum{}B^\dagger_mB^\dagger_n\left\{
	\xi\left(\begin{smallmatrix}n&i_2\\m&i_1\end{smallmatrix}\right)B^{\dagger}_{i_3}\right.\\
	&\left.	+\xi\left(\begin{smallmatrix}n&i_3\\m&i_2\end{smallmatrix}\right)B^{\dagger}_{i_1}	+\xi\left(\begin{smallmatrix}n&i_1\\m&i_3\end{smallmatrix}\right)B^{\dagger}_{i_2}\right\}\left|0\right>
\end{split}
\end{equation}
As in Eq.(\ref{eq:vThree2}), one out of the three excitons seems to stay outside in the scattering process.

Here comes the crucial difference between Wannier excitons and Cooper pairs. Wannier exciton, made of $a^\dagger_{\vk_1}b^\dagger_{\vk_2}$ pairs, have two degrees of freedom. As a direct consequence,  two electrons and two holes can be associated in two different ways to form two excitons. It is possible to show \cite{CobosonPhysicsReports} that
\begin{equation}\label{eq:BB}
B^\dagger_iB^\dagger_j=-\sum\lambda\left(\begin{smallmatrix}n&j\\m&i\end{smallmatrix}%
\right)  B^{\dagger}_{m}B^{\dagger}_{n}
\end{equation}
where $\lambda\left(\begin{smallmatrix}n&j\\m&i\end{smallmatrix}\right)$  is the Pauli scattering of two excitons, defined, as for $\beta^\dagger_\vk$ pairs, through
\begin{equation}
\begin{split}
\left[B_m,B^{\dagger}_{i}\right]&=\delta_{mi}-D_{mi}\\
\left[D_{mi},B^{\dagger}_{j}\right]&=\sum\left\{\lambda\left(\begin{smallmatrix}n&j\\m&i\end{smallmatrix}\right)
+\lambda\left(\begin{smallmatrix}m&j\\n&i\end{smallmatrix}\right)\right\} B^{\dagger}_{n}
\end{split}
\end{equation}

\begin{figure}[htb]
   \includegraphics[width=0.4\textwidth]{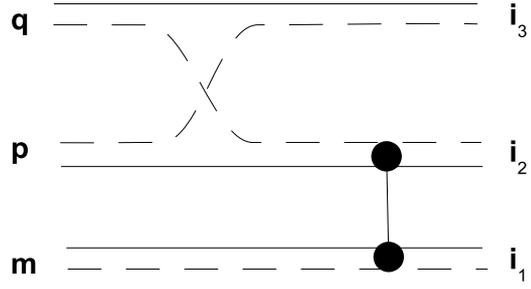}
\caption{Shiva diagram for energy-like exchange interaction between three excitons\label{fig:threeExciton} }
 \end{figure}

The fact that, in the first term of the RHS of Eq.(67), the $B^\dagger_{i_3}$ exciton does not participate to the scattering, is in fact somewhat artificial because, using Eq.(\ref{eq:BB}), we could as well write this first term as
\begin{equation}\label{eq:3BLambdaXi}
\sum{}B^\dagger_mB^\dagger_pB^\dagger_q\left\{\lambda\left(\begin{smallmatrix}q&n\\p&i_3\end{smallmatrix}\right)
+\lambda\left(\begin{smallmatrix}p&n\\q&i_3\end{smallmatrix}\right)\right\} \xi\left(\begin{smallmatrix}n&i_3\\m&i_1\end{smallmatrix}\right)
\end{equation}
This shows that the $(i_1,i_2,i_3)$ excitons can actually be involved in $3\times3$ connected diagram, as shown in Fig. \ref{fig:threeExciton}.

Let us return to the $\beta^\dagger_\vk$ pairs making Cooper pairs. We see that, since the $\vk$ electron with up spin is associated to the $-\vk$ electron with down spin only, an equation similar to Eq.(\ref{eq:BB}) does not exist.  As a result the $\vk_3$ pair in Eq.(\ref{eq:vThree2}) cannot be mixed with the $\vk'_2$ pair as in Eq.(\ref{eq:3BLambdaXi}) to generate $3\times3$ connected diagram, as the one of Fig. \ref{fig:threeExciton}.

\subsection{Richardson's exact eigenstate versus BCS ansatz}

The Richardson's procedure we have here rederived, gives the 
\emph{exact} form of the $H_0+V_{BCS}$ eigenstates as 
\begin{equation}
B^{\dagger}(R_1)\cdots{}B^{\dagger}(R_N)\left|F_0\right>  
\end{equation}
with $B^{\dagger}(R)$ given by Eq.\eqref{eq:B}. The fact that, by
construction, all the $R_i$'s are different in order for the $1/(R_i-R_j)$ factors in the Richardson's equations not to diverge, strongly questions the standard
BCS ansatz for the $N$-pair wave function since this ansatz reduces to $\left(B^{\dagger}\right)
^N\left|F_0\right> $ when projected into the $N$-pair subspace. In this ansatz,  \emph{all} the pairs are taken as condensed into the same state. This is physically hard to accept for composite bosons due to Pauli blocking between pairs which makes each added pair necessarily different from the previous ones, due to the fact that more and more states are occupied already. 

There were in past several discussions about differences and similarities between BCS ansatz and Richardson's exact solution, or more generally a Bethe ansatz like $\prod_iB^\dagger_i\left|0\right>$ from various perspectives and physical situation \cite{bang,hasegawa}.  However, the discussions essentially focus on recovering the correct energy or some other physical quantities, not the wave function itself, more difficult to experimentally evidence. This wave function actually is attached to the picture people commonly have of superconductivity. This is why a correct wave function is importance for physical understanding, at least.

To discuss this problem, let us again start with two pairs. In
a previous work\cite{combescotBCS}, we have shown, that the Richardson's parameters for two pairs read as $R_1=R+iR^{\prime}$ and $R_2=R-i{}R^{\prime}$ with $R$
and $R^{\prime}$ real. In the
large sample limit, i.e. for $1/\rho_0$ small,  the dominant terms of $R$ and $R'$ are given by $R\approx\epsilon_c+1/%
\rho_0+\epsilon_c/\rho_0\Omega$ and $R^{\prime}\approx\sqrt{2\epsilon_c/\rho_0}$. By writing $B^{\dagger}(R_1)$ as $\left[B^{\dagger}(R)+B^{\dagger}(R_1)-B^{\dagger}(R)\right]$ and similarly for  $B^{\dagger}(R_2)$, 
we get, from eq \eqref{eq:B}, 
\begin{equation}\label{eq:BBB2}
B^{\dagger}(R_1)B^{\dagger}(R_2)-\left[B^{\dagger}(R)\right]
^2={R^{\prime}}^2\left\{C^{\dagger}_+C^{\dagger}_--2B^{\dagger}(R)D^{\dagger}%
\right\} 
\end{equation}
where we have set 
\begin{align}
C^{\dagger}_{\pm}&=\sum\frac{w_\vk}{\left(2\epsilon_\vk-R\right)
\left(2\epsilon_\vk-R\pm{}iR^{\prime}\right) }\beta^{\dagger}_\vk \\
D^{\dagger}&=\sum\frac{w_\vk}{\left(2\epsilon_\vk-R\right) \left[%
\left(2\epsilon_\vk-R\right) ^2+{}{R^{\prime}}^2\right] }\beta^{\dagger}_\vk
\end{align}

Eq.(\ref{eq:BBB2}) shows that, in order to possibly replace \\$B^{\dagger}(R_1)B^{\dagger}(R_2)$ by $\left[B^{\dagger}(R)\right]
^2$ as in the BCS ansatz, we must neglect terms in $R'^2$, i.e., in $1/\rho_0$. However, these $1/\rho_0$ terms are precisely those which make $E _1$ different from $E _2/2\approx E_1+1/\rho_0+\epsilon_c/\rho_0\Omega$; so that the replacement of $B^{\dagger}(R_1)B^{%
\dagger}(R_2)$ by a ``condensed two-pair state'' $\left(B^{\dagger}(E _2/2)\right) ^2$ with $E_2$ \emph{different }from $E_1$ is fully inconsistent because, in this two-pair
operator, we would keep contributions in $1/\rho_0$ which are as large as the ones we drop
by neglecting the RHS of Eq.(\ref{eq:BBB2}) : two pairs do not condense into the same state.

Actually, it is claimed that the BCS ansatz is valid in the thermodynamical
limit when $N$ is very large. It is possible to show that, for $N$ large but still in the dilute regime on the single pair scale, the $R_i$'s stay two by two complex conjugate, the imaginary part of $R_i $'s getting larger and larger as $\sqrt{N\epsilon_c/\rho_0}$ when $N$ increases. By using a similar procedure as the one we used for $N=2$, we hardly see how, 
starting from the exact form of the $N$-pair eigenstate $B^{\dagger}(R_1)\cdots{}B^{\dagger}(R_N)\left|F_0\right>  $, we can possibly recover the BCS ansatz with the \emph{same} 
creation operator for all  pairs when $N$ ventures outside the dilute limit because nothing special happens in the behavior of the $R_i$'s when $N$ crosses $N_c$. 

In a recent work, Ortiz and Dukelsky\cite{crossoverRich} have also considered Richardson's equations in the thermodynamical limit. While they do recover the energy obtained from the ansatz, they conclude, like us, that Richardson's exact wave function is substantially different from the BCS ansatz in many ways.

We wish to stress that, 
to the best of our knowledge,  derivations of the validity of the BCS ansatz for the ground state of $N$ pairs
mainly concentrate on the energy it provides 
(see, e.g., \cite{Schrieffer} and references therein).
 We of course agree that the BCS
ansatz gives the correct ground state energy for $N$ pairs because the energy obtained using this ansatz
is just the one we derived from the exact Richardson's procedure, extrapolated outside the dilute limit. However,
agreement on the energy by no mean proves agreement on the wave function.
Many examples have been given in the past with wave functions very different
from the exact one, although giving correct energy. Direct
experiments supporting the form of the ground state wave function however seems to be even harder to achieve than the ones possibly checking the $N$ dependence of the ground state energy given in Eq.(1). 
Nevertheless, it seems to us highly desirable to carefully reconsider ``agreement 
with experiments'' in the light of the exact Richardson's wave function. It is still a rather intriguing question to understand why the minimization of the hamiltonian mean value calculated with this ansatz, leads to \emph{exactly} the same energy as the one we derived by analytically solving Richardson's equations in the dilute limit.

It is worth noting that the reduced potential used in standard BCS superconductivity has the great advantage to allow an analytical resolution of the $N$-body Schr\"{o}dinger equation - which is quite infrequent. It however is clear that this potential is highly simplified. A certain amount of corrections are necessary to make this potential more realistic. These are going to destroy the possibility to get the eigenstates analytically. However, since the BCS ansatz for the wave function with all the pairs condensed into the \emph{same} state - which is commonly considered as one of the essential features of superconductivity - has been worked out within this reduced BCS potential, a precise comparison between this conventional ansatz and the exact solution of the model in the canonical ensemble, is definitely quite relevant to better understand the deep physics hidden in this ansatz.

Finally, we wish to stress that the possible replacement of $B^{\dagger}(R_1)\cdots{}B^{\dagger}(R_N)%
\left|F_0\right>  $ by $\left(B^{\dagger}\right) ^N\left|F_0\right>  $ is
crucial to support the overall picture of
superconductivity commonly in mind, with all the pairs in the same state, ``as an army of
little soldiers, all walking similarly''.  This picture actually seems a rather naive extrapolation to composite bosons, of the standard Bose-Einstein condensation demonstrated in the case of  \emph{elementary} bosons. It is hard for us to accept that, in the case of composite bosons, Pauli blocking between fermionic components is not going to destroy nice harmony in this ``army''.
More work on the validity of the BCS 
ansatz in the thermodynamical limit in the context of the coboson nature of Cooper pairs, seems a necessity to more deeply understand some unrevealed aspects of basic superconductivity as the ones at the origin of Eq.(1) for the $N$-pair ground state energy. The coboson many-body formalism we have here constructed, should appear as quite valuable because it gives a fresh view to this famous field, its Shiva diagram representation helping to support physical understanding.

\section{Conclusion}

We have constructed a coboson formalism for the electron pairs on which Cooper pairs are made. It has similarity with the one we have constructed for composite boson excitons. This formalism evidences that the scatterings of two zero-momentum electron pairs in the BCS potential, are mediated by the Pauli exclusion principle. No direct process exists

As a first application, we here rederive Richardson's equations for the exact eigenstates of $N$ Cooper pairs. This 
  derivation allows us to trace back
the physical origin of the various terms. In
particular, we clearly see that $N$ pairs differ from $N$ independent
pairs, due to Pauli blocking only. This Pauli blocking
enforces the $R_i$ parameters of Richardson's equations to be all different. 
As a direct consequence, the exact wave function for $N$ interacting pairs is 
definitely different from the BCS ansatz, although the $N$-pair energy this ansatz provides, is the correct one in the large sample limit. 

The diagrammatic
representation of our derivation also shows that, because electron pairs
with zero total momentum have one degree of freedom instead of two, they scatter within the BCS potential through 
$(2\times2)$ exchange interaction scatterings only. This possibly explains
why the $N$-pair ground state energy that we have recently found in the dilute limit, has interaction terms in $N(N-1)$ but not in $N(N-1)(N-2)
$ and higher, as expected for $N$-body problems.

One of us (M.C.) wishes to thank the University of Illinois at
Urbana-Champaign, and Tony Leggett in particular, for enlightening discussions during her invitation at
the Institute for Condensed Matter Physics where most of the present work has been
performed. We also wish to thank Walter Pogosov for his constructive comments on the manuscript.

\bibliographystyle{epj}
\bibliography{citation}
\end{document}